\documentclass[iop]{emulateapj}

\newcommand{\QSO}     {SDSS\,J2222$+$2745}
\newcommand{\QSOLone}    {SDSS\,J1004$+$4112}
\newcommand{\QSOLtwo}    {SDSS\,J1029$+$2623}
\newcommand{\etal}    {{\it et al.~}}

\newcommand{\magA}{$5.7 ^{+14.5}_{-3.1}$}
\newcommand{\magB}{$5.6 ^{+12.8}_{-2.2}$}
\newcommand{\magC}{$2.5 ^{+3.6}_{-0.6}$}
\newcommand{\magD}{$0.6 ^{+0.7}_{-0.2}$}
\newcommand{\magE}{$0.7 ^{+0.9}_{-0.3}$}

\newcommand{\magAsix}{$4.7 ^{+3.4}_{-1.6}$}
\newcommand{\magBsix}{$5.4 ^{+4.6}_{-1.6}$}
\newcommand{\magCsix}{$2.4 ^{+1.1}_{-0.3}$}
\newcommand{\magDsix}{$0.7 ^{+0.3}_{-0.2}$}
\newcommand{\magEsix}{$0.6 ^{+0.3}_{-0.2}$}
\newcommand{\magFsix}{$0.5 ^{+0.5}_{-0.2}$}

\newcommand{\timeA}{$  0 $}
\newcommand{\timeB}{$-87 ^{+187}_{-296}$}
\newcommand{\timeC}{$-1399 ^{+850}_{-776}$}
\newcommand{\timeD}{$960 ^{+365}_{-254}$}
\newcommand{\timeE}{$851 ^{+448}_{-278}$}

\newcommand{\timeAsix}{$  0$}
\newcommand{\timeBsix}{$-112 ^{+158}_{-225}$}
\newcommand{\timeCsix}{$-1368 ^{+456}_{-344}$}
\newcommand{\timeDsix}{$931 ^{+237}_{-233}$}
\newcommand{\timeEsix}{$806 ^{+370}_{-224}$}
\newcommand{\timeFsix}{$723 ^{+223}_{-161}$}

\newcommand{\Px}{$-0.14 ^{+0.50}_{-1.03}$}
\newcommand{\Py}{$1.18 ^{+1.90}_{-0.81}$}
\newcommand{\Pe}{$0.66 ^{+0.25}_{-0.14}$}
\newcommand{\Ptheta}{$ 97 ^{+  5}_{-  4}$}
\newcommand{\Prc}{$ 26 ^{+ 15}_{- 42}$}
\newcommand{\Psigma}{$580 ^{+ 64}_{-163}$}
\newcommand{\PrcA}{$0.77 ^{+0.77}_{-0.23}$}
\newcommand{\PsigmaA}{$155 ^{+ 55}_{-168}$}
\newcommand{\PrcB}{0.13 [0.0 0.3]}
\newcommand{\PsigmaB}{$212 ^{+ 40}_{- 44}$}
\newcommand{\PrcC}{0.19 [0.0 0.3]}
\newcommand{\PsigmaC}{$279 ^{+ 76}_{- 21}$}

\newcommand{\Pxsix}{$-0.16 ^{+0.58}_{-0.71}$}
\newcommand{\Pysix}{$2.10 ^{+2.58}_{-0.90}$}
\newcommand{\Pesix}{$0.75 ^{+0.13}_{-0.05}$}
\newcommand{\Pthetasix}{$ 96 ^{+  4}_{-  3}$}
\newcommand{\Prcsix}{$ 37 ^{+ 13}_{- 40}$}
\newcommand{\Psigmasix}{$578 ^{+ 50}_{-114}$}
\newcommand{\PrcAsix}{$0.05 ^{+0.04}_{-0.93}$}
\newcommand{\PsigmaAsix}{$303 ^{+ 34}_{- 46}$}
\newcommand{\PrcBsix}{0.15 [0.0 0.3]}
\newcommand{\PsigmaBsix}{$226 ^{+ 29}_{- 36}$}
\newcommand{\PrcCsix}{0.15 [0.0 0.3]}
\newcommand{\PsigmaCsix}{$235 ^{+ 46}_{- 33}$}

\shorttitle{\QSO : 15\farcs 1 sextuple QSO}
\slugcomment{ApJ in prep: draft date \today}
\shortauthors{Dahle \etal}

\begin{document}
\title{\QSO : A Gravitationally Lensed Sextuple Quasar with Maximum Image Separation of 15\farcs 1 Discovered in the Sloan Giant Arcs Survey. \footnote{Based on observations made with the Nordic Optical Telescope, operated
on the island of La Palma jointly by Denmark, Finland, Iceland, Norway, and Sweden, in the Spanish Observatorio del Roque de los Muchachos of the Instituto de Astrofisica de Canarias.   }}
\author{
H.~Dahle\altaffilmark{1,2}, M.~D.~Gladders\altaffilmark{3,4}, K.~Sharon\altaffilmark{5}, M.~B.~Bayliss\altaffilmark{6,7}, E.~Wuyts\altaffilmark{8}, L.~E.~Abramson\altaffilmark{3,4}, B.~P.~Koester\altaffilmark{9}, N.~Groeneboom\altaffilmark{1}, T.~E.~Brinckmann\altaffilmark{10}, M.~T.~Kristensen\altaffilmark{10}, M.~O.~Lindholmer\altaffilmark{10}, A.~Nielsen\altaffilmark{10}, J.-K.~Krogager\altaffilmark{10}, J.~P.~U.~Fynbo\altaffilmark{10}} 

\email{hdahle@astro.uio.no} 

\altaffiltext{1} {Institute of Theoretical Astrophysics, University of Oslo, P. O. Box 1029, Blindern, N-0315 Oslo, Norway} 
\altaffiltext{2} {Centre of Mathematics for Applications, University of Oslo, Blindern, Oslo, Norway}
\altaffiltext{3} {Department of Astronomy \& Astrophysics, The University of Chicago, 5640 S.\ Ellis Avenue, Chicago, IL 60637, USA} 
\altaffiltext{4} {Kavli Institute for Cosmological Physics at the University of Chicago}
\altaffiltext{5} {Department of Astronomy, University of Michigan, 500 Church Street, Ann Arbor, MI 48109, USA} 
\altaffiltext{6} {Harvard-Smithsonian Center for Astrophysics, 60 Garden Street, Cambridge, MA 02138, USA} 
\altaffiltext{7} {Department of Physics, Harvard University, 17 Oxford Street, Cambridge, MA 02138} 
\altaffiltext{8} {Max-Planck-Institut f{\"u}r extraterrestrische Physik, Giessenbackstrasse 1, 85748 Garching bei M{\"u}nchen, Germany} 
\altaffiltext{9} {Department of Physics, University of Michigan, 450 Church Street, Ann Arbor, MI 48109, USA} 
\altaffiltext{10} {Dark Cosmology Centre, Niels Bohr Institute, Juliane Maries Vej 30, DK-2100 Copenhagen O, Denmark}

\begin{abstract}
We report the discovery of a unique gravitational lens system, \QSO , producing five spectroscopically confirmed images of a $z_s = 2.82$ 
quasar lensed by a foreground galaxy cluster at $z_l = 0.49$.
We also present photometric and spectroscopic evidence for a sixth lensed image of the same quasar. 
The maximum separation between the quasar images is $15\farcs 1$. Both the large image separations and the high image 
multiplicity of the lensed quasar are in themselves rare, 
and observing the combination of these two factors is an exceptionally unlikely occurrence in present datasets. 
This is only the third known case of a quasar lensed by a cluster, and the only one with six images.  
The lens system was discovered in the course of the Sloan Giant Arcs Survey, in which we identify candidate lenses in the 
Sloan Digital Sky Survey and target these for follow up and verification with the 2.56m Nordic Optical Telescope.  
Multi-band photometry obtained over multiple epochs from September 2011 to September 2012 reveals significant variability at the $\sim 10-30\%$ level in some of the quasar images, 
indicating that measurements of the relative time delay between quasar images will be feasible.  
In this lens system we also identify a bright ($g = 21.5$) giant arc corresponding to a strongly lensed background galaxy at $z_s = 2.30$. We fit parametric models of the lens system, constrained by the redshift and positions of the quasar images and the redshift and position of the giant arc. The predicted time delays between different pairs of quasar images range from $\sim 100$\,days to $\sim 6$\,years. 

\end{abstract}


\section{Introduction} 
\label{sec:intro}

Strongly gravitationally lensed quasars with image separations $\gtrsim 10"$ are exceptionally rare. Such large image splittings 
require a lensing mass corresponding to a galaxy cluster, and a fortuitous close line-of-sight alignment of such a mass distribution and a background
quasar. In fact, only two such systems have previously been discovered, \QSOLone \, (Inada et al.\ 2003) with a maximum image separation of $14\farcs 6$ and \QSOLtwo \, (Inada et al.\ 2006)   
with a maximum image separation of $22\farcs 5$. Such systems provide unique information on the abundances and internal structure of galaxy clusters, yielding stringent tests of the 
current $\Lambda$CDM paradigm, both through their observed abundances (e.g., Hennawi et al.\ 2007), the relative fractions of different image multiplicities (e.g., Oguri \& Keeton 2004), and from detailed models  
of the cluster mass distribution based on the available constraints (e.g., Sharon et al.\ 2005; Oguri 2010; Oguri et al.\ 2012), 
including measured time delays between the quasar images (Fohlmeister et al.\ 2008, 2012). Being effectively point sources, the positions of multiply lensed quasar images provide mass modelling constraints which are free from the ambiguities which sometimes plague the identification of multiple images of strongly lensed galaxies, particularly the fainter de-magnified images located closest to the mass center.  

As shown by Refsdal (1964), the relative time delays of multiple quasar images visible around the lensing mass center can provide an estimate of the Hubble parameter $H_0$. However, 
as $H_0$ is presently increasingly well constrained by a host of methods, including gravitational lensing time delays from galaxy-scale lenses, the main interest of quasar time delays are 
related to their ability to provide unique detailed constraints on the slope of the projected matter density distribution over the range of radii where the quasar images are found (e.g., Refsdal 2004).    
    
In this paper, we present the discovery of a third lensing system
where a quasar is being multiply lensed by a galaxy cluster, but with
a higher image multiplicity than the previously known such
systems. The discussion in subsequent sections of this paper describes
the evidence for the lensing interpretation of this system and the
identification of an increasing number of components in the lensing
system, through analysis of progressively deeper photometric and
spectroscopic data, coupled with parametric modelling of the lens.
 
In \S~\ref{sec:ima1} we describe the initial identification of \QSO \,
as a likely gravitational lens system in the course of a larger survey
for strong cluster lenses.  The spectroscopic confirmation of the
three brightest lensed images of the quasar is outlined in
\S~\ref{sec:ABCspec}, and \S~\ref{sec:DEF} details the evidence for
three additional fainter quasar images from deeper imaging of the
field, coupled with further spectroscopic observations. In
\S~\ref{sec:phot} we report photometry of the quasar images, some
showing significant brightness changes over a time span of one
year. In \S~\ref{sec:cluster} we provide a summary of the observed
properties of the lensing cluster, and in \S~\ref{sec:model} we
develop a parametric model of the lens system, which is used to
predict time delays. Our main results are summarized in
\S~\ref{sec:discuss}.  Unless otherwise noted, throughout this work we
assume an $\Omega_m = 0.3$, $\Omega_{\Lambda} = 0.7$, $H_0 = 70$ km
s$^{-1}$ cosmology.

\section{Lens candidate selection and initial imaging} 
\label{sec:ima1}

We have undertaken a search for cluster lenses, the Sloan Giant Arcs Survey (SGAS; Hennawi et al.\ 2008; Bayliss et al.\ 2011; Gladders et al.\ in prep.\ 2012), based on public data from the Sloan Digital Sky Survey (SDSS; York et al.\ 2000). 
Potential lines of sight for lensing by clusters are identified by running an automated cluster finding algorithm on the SDSS imaging data. This is followed by visual inspection, 
by multiple persons in our collaboration, of color images of these lines of sight, generated from the SDSS data. New lens candidates are flagged and given a grade according to the likelihood of 
being a true lens, and a final grade is calculated for each lens candidate. The list of lens candidates is then checked against public data bases such as the NASA/IPAC Extragalactic Database, 
and remaining candidates after removal of known lenses are targeted for deeper follow-up imaging. In September 2011, we targeted such lens candidates selected from the SDSS Data Release 8 (Aihara et al.\ 2011) 
using the MOsaic CAmera (MOSCA) at the 2.56m Nordic Optical Telescope (NOT). MOSCA is a $2\times 2$ mosaic of four $2048 \times 2048$ Loral-Lesser thinned CCDs with excellent blue sensitivity 
(Quantum Efficiency $> 90\%$ across the SDSS $g$-band), mounted at the Cassegrain focus of the telescope. MOSCA is normally used in $2\times 2$ binned mode, yielding a pixel scale 
of $0.217\arcsec {\rm pixel}^{-1}$ and a total field of view of $7\farcm 7 \times 7\farcm 7$.    

\begin{deluxetable*}{lccc}
\tabletypesize{\footnotesize}
\tablecaption{ Log of NOT/MOSCA imaging observations\label{tab:MOSCAlog}}
\tablehead{
\colhead{Epoch } & \colhead{Filter}    & \colhead{Exposure time}  & \colhead{FWHM}
}
\startdata
\hline \\                     
{\bf 2011 Sep.\ 24.93 UT} & $g$ & $2\times 300$s & $0\farcs 87$ \\
{\bf 2011 Sep.\ 24.94 UT} & $r$ & $2\times 150$s & $0\farcs 81$ \\
{\bf 2011 Sep.\ 24.95 UT} & $i$ & $2\times 150$s & $0\farcs 74$ \\
\hline \\
{\bf 2012 Sep.\ 13.97 UT} & $u$ & $3\times 600$s & $0\farcs 81$ \\
{\bf 2012 Sep.\ 12.96 UT} & $g$ & $3\times 600$s & $0\farcs 70$ \\
{\bf 2012 Sep.\ 13.04 UT} & $r$ & $3\times 300$s & $0\farcs 68$ \\
{\bf 2012 Sep.\ 12.95 UT} & $i$ & $3\times 300$s & $0\farcs 61$ \\
\hline \\
{\bf 2012 Sep.\ 15.98 UT} & $g$ & $3\times 600$s & $0\farcs 68$ \\
\end{deluxetable*}

\begin{figure*}
\begin{center}
\includegraphics[angle=0,scale=.60]{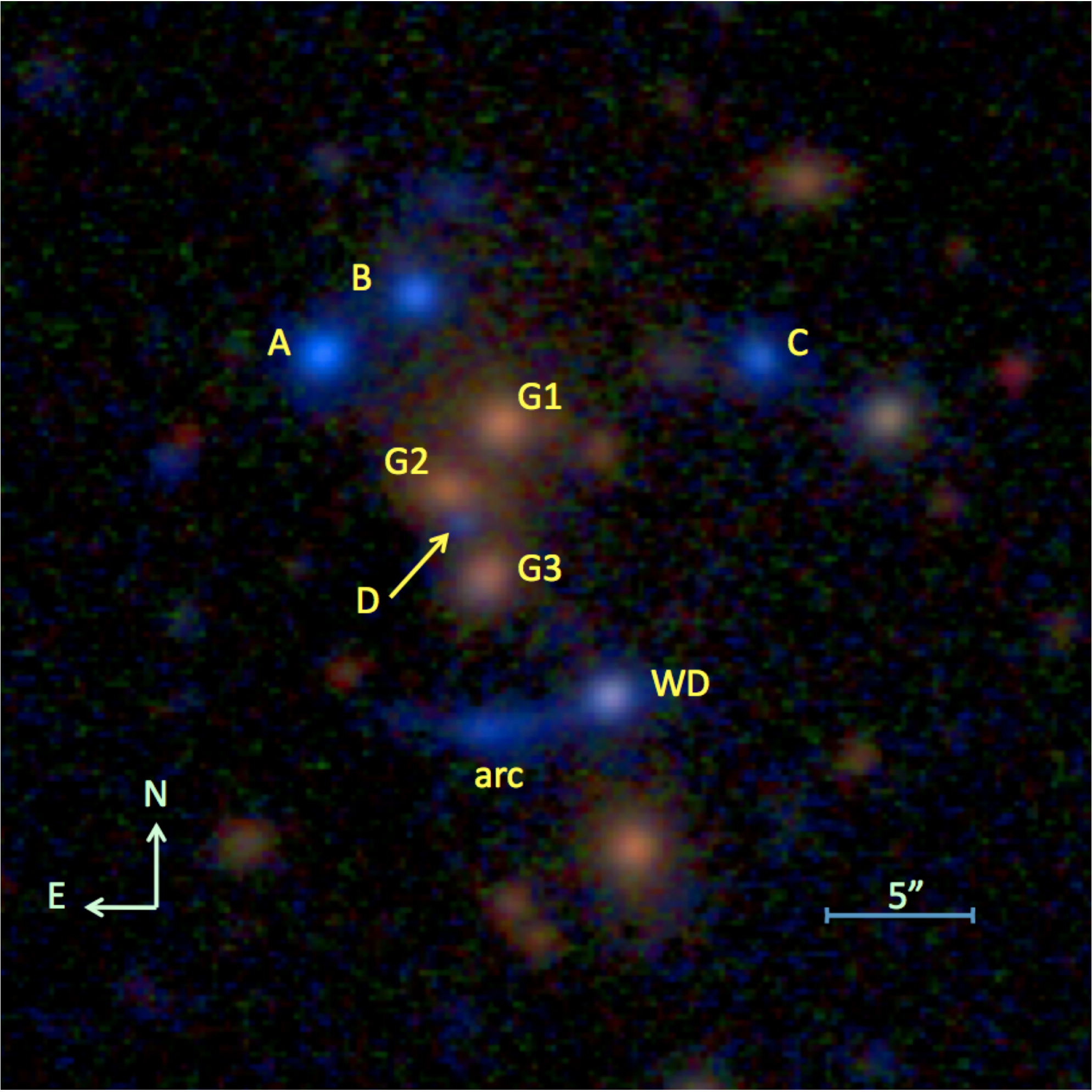}
\end{center}
\caption[NOT/ALFOSC spectra of the three brightest quasar images.]
{NOT/MOSCA $gri$ color composite image of \QSO \, from our initial epoch of photometry (see Table~\ref{tab:MOSCAlog}).}
\label{fig:Firstepoch}
\end{figure*}

\QSO \, was imaged using MOSCA in September 2011 (see Table~\ref{tab:MOSCAlog} for details) 
with our nominal survey exposure time of $2\times 300$s in the $g$-band. Based on a visual inspection of the raw data, it was noted as a probable lens, 
and subject to immediate follow-up with $2\times 150$s exposures in each of the SDSS $r$- and $i$-bands. In addition to the apparent blue arc south of the cluster core, 
three blue stellar images (labelled ``A'',''B'', and ''C'' in Figure~\ref{fig:Firstepoch}) located at similar cluster-centric radii as the arc, 
were considered as candidate multiple images of a gravitationally lensed quasar. 
In addition, three red galaxies, 
labelled ``G1-G3'' in Figure~\ref{fig:Firstepoch}, were seen in the cluster core. 
Publicly available photometric redshift estimates from SDSS (see Table~\ref{tab:astrom}) indicated that all three galaxies are consistent with being located in a cluster at $z\sim 0.50$.
    
The stellar source (labelled ``WD'' in Figure~\ref{fig:Firstepoch}) of magnitude $g=21.09$, similar in apparent brightness to A and B, seen superposed on the giant arc, was also noted as a candidate quasar image. 
Although this image has a significantly redder color ($g-i = 1.33$) than A and B, it might be considerably reddened by extinction in galaxies (the galaxy corresponding to the arc being one potential 
such candidate) along its line of sight, and it was kept as a target for future spectroscopic follow-up.   
    
\begin{deluxetable*}{lcccccccc}
\tabletypesize{\footnotesize}
\tablecaption{ SDSS DR8 astrometry, photometry, and photometric redshift estimates for the three brightest quasar images and three brightest galaxies in \QSO \label{tab:astrom}}
\tablehead{
\colhead{Object} & \colhead{RA}    & \colhead{Dec}  & \colhead{u} & \colhead{g} & \colhead{r} & \colhead{i} & \colhead{z} &  \colhead{Photometric} \\
      & \colhead{(h m s)} & \colhead{(d m s)}  & \colhead{SDSS} &  \colhead{SDSS} & \colhead{SDSS} & \colhead{SDSS} & \colhead{SDSS} &  \colhead{redshift} 
}
\startdata
\hline \\                     
A            & 22:22:09.04 & +27:45:37.9 & 24.55$\pm$0.93 & 21.10$\pm$0.03 & 20.67$\pm$0.04 & 20.87$\pm$0.07 & 21.00$\pm$0.27 & \nodata \\
B            & 22:22:08.80 & +27:45:40.0 & 22.07$\pm$0.19 & 21.43$\pm$0.04 & 21.03$\pm$0.05 & 21.27$\pm$0.10 & 21.08$\pm$0.29 & \nodata \\
C            & 22:22:07.90 & +27:45:37.8 & 24.50$\pm$0.93 & 21.51$\pm$0.05 & 21.14$\pm$0.05 & 21.28$\pm$0.10 & 22.14$\pm$0.63 & \nodata \\
G1            & 22:22:08.57 & +27:45:35.6 & 25.42$\pm$2.41 & 20.92$\pm$0.08 & 19.00$\pm$0.03 & 18.05$\pm$0.02 & 17.65$\pm$0.05 & 0.485$\pm$0.015 \\
G2            & 22:22:08.70 & +27:45:33.0 & 24.33$\pm$1.01 & 22.13$\pm$0.08 & 20.86$\pm$0.05 & 19.88$\pm$0.04 & 19.33$\pm$0.07 & 0.528$\pm$0.157 \\
G3            & 22:22:08.62 & +27:45:29.9 & 24.39$\pm$2.32 & 22.28$\pm$0.20 & 20.43$\pm$0.07 & 19.48$\pm$0.05 & 19.13$\pm$0.16 & 0.509$\pm$0.031 \\
\tablecomments{Epoch: 2009 October 18.19 UT}
\end{deluxetable*}

\section{Follow-up observations and results} 
\label{sec:results}

\subsection{Spectroscopic confirmation of the three brightest quasar images} 
\label{sec:ABCspec}

The spectroscopic observations described in this paper were obtained using the Andalucia Faint Object Spectrograph and Camera (ALFOSC) at the NOT. We used grism \# 4, covering a wavelength range 
$3800 {\rm {\AA}} - 9000 {\rm {\AA}}$ with a resolution of about $R=350$ and dispersion 3.0 {\AA} pixel$^{-1}$. The ALFOSC pixels were binned by a factor of two in the dispersion direction
but kept unbinned at the native pixel scale of 0\farcs 188 pixel$^{-1}$ in the spatial direction. The slit width was 1\farcs 3. Redwards of $\sim 7000$ {\AA} strong fringing is present in the spectra, effectively 
restricting the upper limit of the spectral range to this value unless special care is taken during the observations to suppress the effects of fringing. An order-sorter filter was used to suppress 
second-order contamination in the spectra. 

\begin{figure}
\plotone{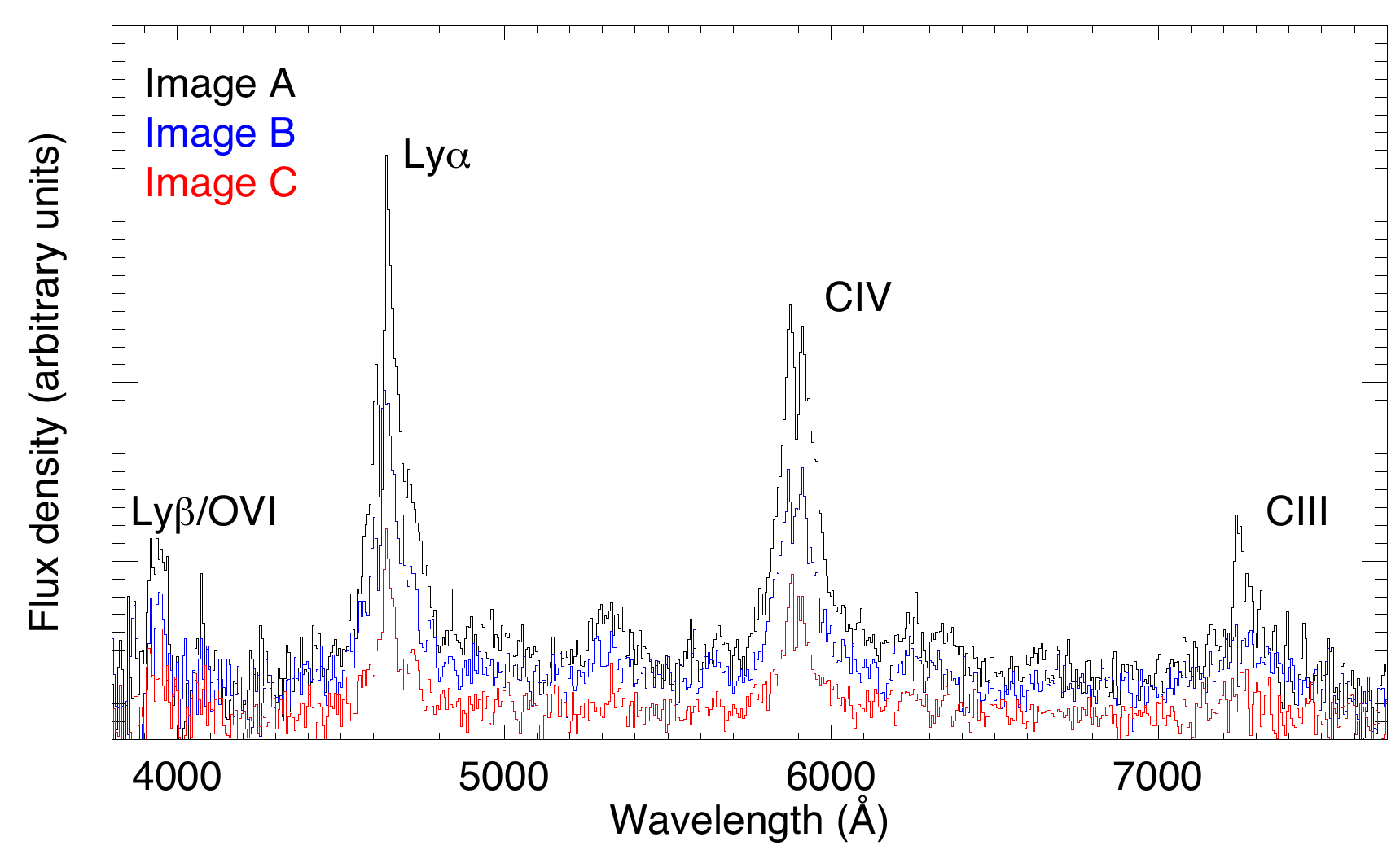}
\caption[NOT/ALFOSC spectra of the three brightest quasar images.]
{NOT/ALFOSC spectra of the three brightest quasar images A-C. The most prominent quasar emission lines, redshifted to $z=2.82$, are indicated in the figure.}
\label{fig:Johanfig}
\end{figure}

On August 21-22, 2012, a series of ALFOSC spectra were obtained in the area of \QSO : Three 1200s spectra were obtained with the rotation angle set to provide simultaneous coverage of 
the A and B images. In addition, two single spectra of 1000s exposure time were obtained to simultaneously cover image C and galaxy G1 and image B and image C, respectively. 
Finally, the blue giant arc and the stellar source superposed on it (labelled ``WD'' in Figure~\ref{fig:Firstepoch}) were observed for a total exposure time $3\times 1000$s. 
The spectra were wavelength calibrated using He$+$Ne arc lamps and flux calibrated based on observations of the spectrophotometric standard star BD$+$33~2642. 
As shown in Figure~\ref{fig:Johanfig} the three images A,B, and C all have identical spectra modulo noise, containing typical spectral features of a quasar (e.g., Vanden Berk et al.\ 2001) with a set of 
prominent emission lines, redshifted to $z=2.82$. This demonstrates that the lensing interpretation is correct for these three quasar images.       

Despite having a significantly redder color, the stellar source superposed on the giant arc was also 
targeted for ALFOSC spectroscopy as a potential additional image of the quasar. Although a significant continuum was detected, emission lines were completely absent at the wavelengths 
of the prominent quasar emission lines in the A,B, and C component spectra. A more careful analysis of the spectrum reveals the typical signatures of a DZ white dwarf (e.g., Koester et al.\ 2011).   

\subsection{Identification of the D, E and F images} 
\label{sec:DEF}

Having confirmed the three images A-C as gravitationally lensed images
of the same quasar, two alternative image configurations may be
considered: \QSO \, is either a naked cusp lens with three quasar
images, similar to \QSOLtwo \, (Oguri et al.\ 2008) or a five-image
system similar to \QSOLone \, (Inada et al.\ 2008). Being able to
distinguish between the two cases is important, not only for the
understanding of the \QSO \, lens system itself, but also in the
context of using the statistics of cluster-lensed quasars as a
cosmological tool. Naked cusps, i.e., lensing configurations where the source lies on a
tangential caustic that extends outside a radial caustic, are extremely
rare among galaxy-scale lenses, the only known candidate among $~\sim
100$ such lens systems being APM~08279$+$5255 (Lewis et
al.\ 2002). However, due to their shallower central matter density
slopes, naked-cusp lenses are expected to be much more common for
clusters than for galaxies.  Oguri \& Keeton (2004) predicted that
$30-60\% $ of all cluster lenses will be of this type, the fraction
depending on the central slope and triaxiality of clusters.

To further investigate the lensing system configuration of \QSO \, we
searched our initial MOSCA imaging for potential stellar sources with
blue colors consistent with the quasar. A blue source, labelled ``D''
in Figure~\ref{fig:Firstepoch}, was identified just south of, and
partially superposed on, galaxy G2. The source was apparently stellar,
but the modest exposure time and overlap with G2 made this conclusion
somewhat uncertain.  A set of deeper $ugri$ images in better seeing
was obtained at the NOT in September 2012 (see
Table~\ref{tab:MOSCAlog}). These images revealed image D to have a
clearly stellar point-spread-function (PSF), motivating further
spectroscopic follow-up of this image. On 2012 September 16-17 UT, a
series of $8 \times 2400$s integrations were made using ALFOSC, using
the same setup as for the earlier observations described in
\S~\ref{sec:ABCspec}.  The telescope pointing was dithered along the
slit between exposures to improve fringe suppression at the red end of
the spectrum. The spectra were flux calibrated based on observations
of the spectrophotometric standard star BD$+$33~2642.

\begin{figure}
\includegraphics[angle=0,scale=.40]{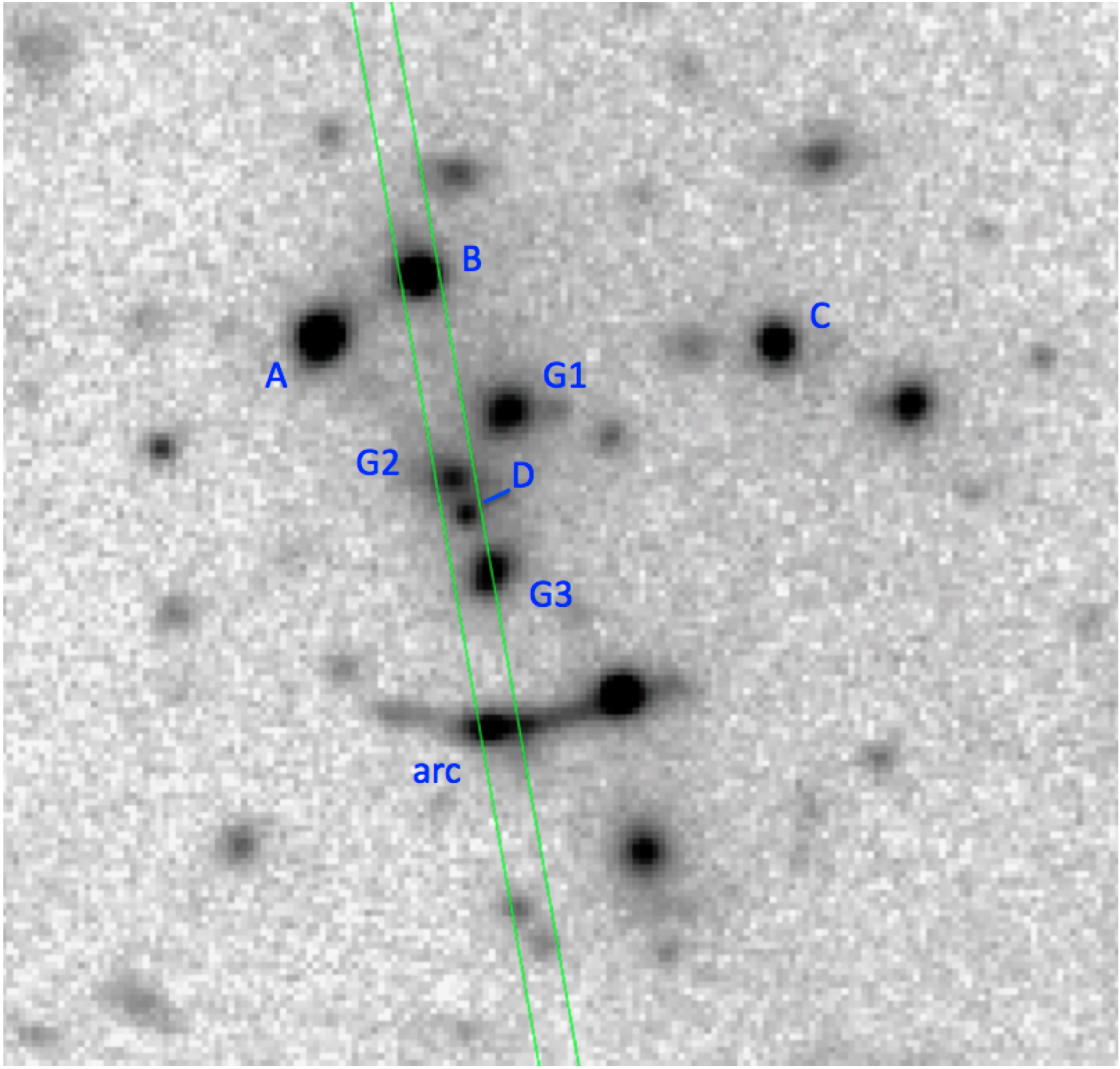}
\caption[.]  {Position of the slit for spectroscopic observations of
  \QSO \, using ALFOSC in September 2012. The field of view and
  orientation is the same as for Figure~\ref{fig:Firstepoch}. The
  background image is a 1h $g$-band MOSCA image combining all
  exposures obtained in this passband in September 2012 (see
  Table~\ref{tab:MOSCAlog}).}
\label{fig:Slitpos}
\end{figure}

\begin{figure}
\includegraphics[angle=0,scale=.35]{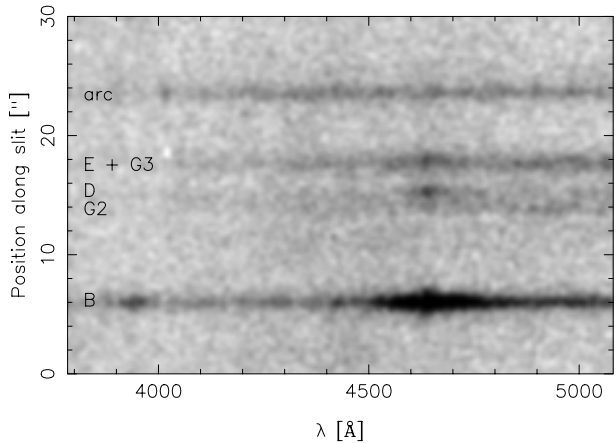}
\caption[.]  {Reduced 2D spectrum with slit orientation as shown in
  Figure~\ref{fig:Slitpos}. The brightest spectrum near the bottom is
  the B image of the quasar. The locations of other sources along the
  slit are indicated along the left margin.  The D and E images (the
  latter superposed on the spectrum of galaxy G3) display clear
  Lyman-alpha emission lines at the same redshift as the B image. The
  image has been gaussian smoothed on a scale of FWHM$=12${\AA} in the
  spectral direction and FWHM$=0.4\arcsec$ in the spatial direction.}
\label{fig:2Dspec}
\end{figure}

\begin{figure}
\includegraphics[angle=0,scale=.35]{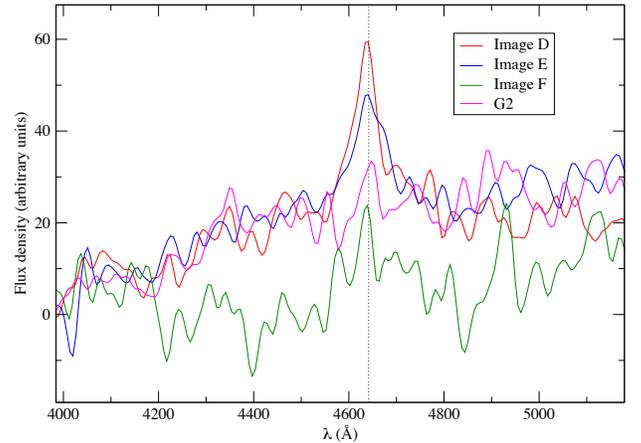}
\caption[.]  {Spectra of the D, E, and F images and galaxy G2, showing
  the spectral region surrounding the Lyman-alpha emission line. The
  wavelength of Lyman-alpha at the $z=2.82$ redshift of the quasar is
  indicated by the vertical dotted line. To dampen small-scale noise
  the spectra have been smoothed with a Gaussian filter of
  FWHM$=18${\AA}.  }
\label{fig:DEFspecs}
\end{figure}

The slit orientation is illustrated in Figure~\ref{fig:Slitpos},
providing simultaneous spectroscopic coverage of images B and D,
galaxies G2 and G3, and the giant arc.  Inspection of the reduced 2D
spectrum (see Figure~\ref{fig:2Dspec}) reveals the strong spectral
line in the quasar B image around 4640~{\AA}, corresponding to
Lyman-alpha at the redshift of the quasar.  A Lyman-alpha line is also
clearly evident at this wavelength in the spectrum of the D image,
confirming its identification as a gravitationally lensed image of the
quasar.  More surprisingly, a Lyman-alpha line at the same redshift is
also seen superposed on the spectrum of galaxy G3, providing clear
indication of a fifth (E) quasar image spatially coincident with this
galaxy. The extracted 1D spectra of image D and E in a $\sim
1200${\AA} wide region surrounding the Lyman-alpha line of the quasar
are plotted in Figure~\ref{fig:DEFspecs}: A clear peak is seen at the
location of the Lyman-alpha line in both the D and E image
spectra. Redward of this line, there is some visible discrepancy in
the slopes of the two spectra. This discrepancy is caused by the
spatial superposition of image E and the red galaxy G3, the latter
having a continuum which is rising towards longer wavelengths.
 
\begin{figure}
\plotone{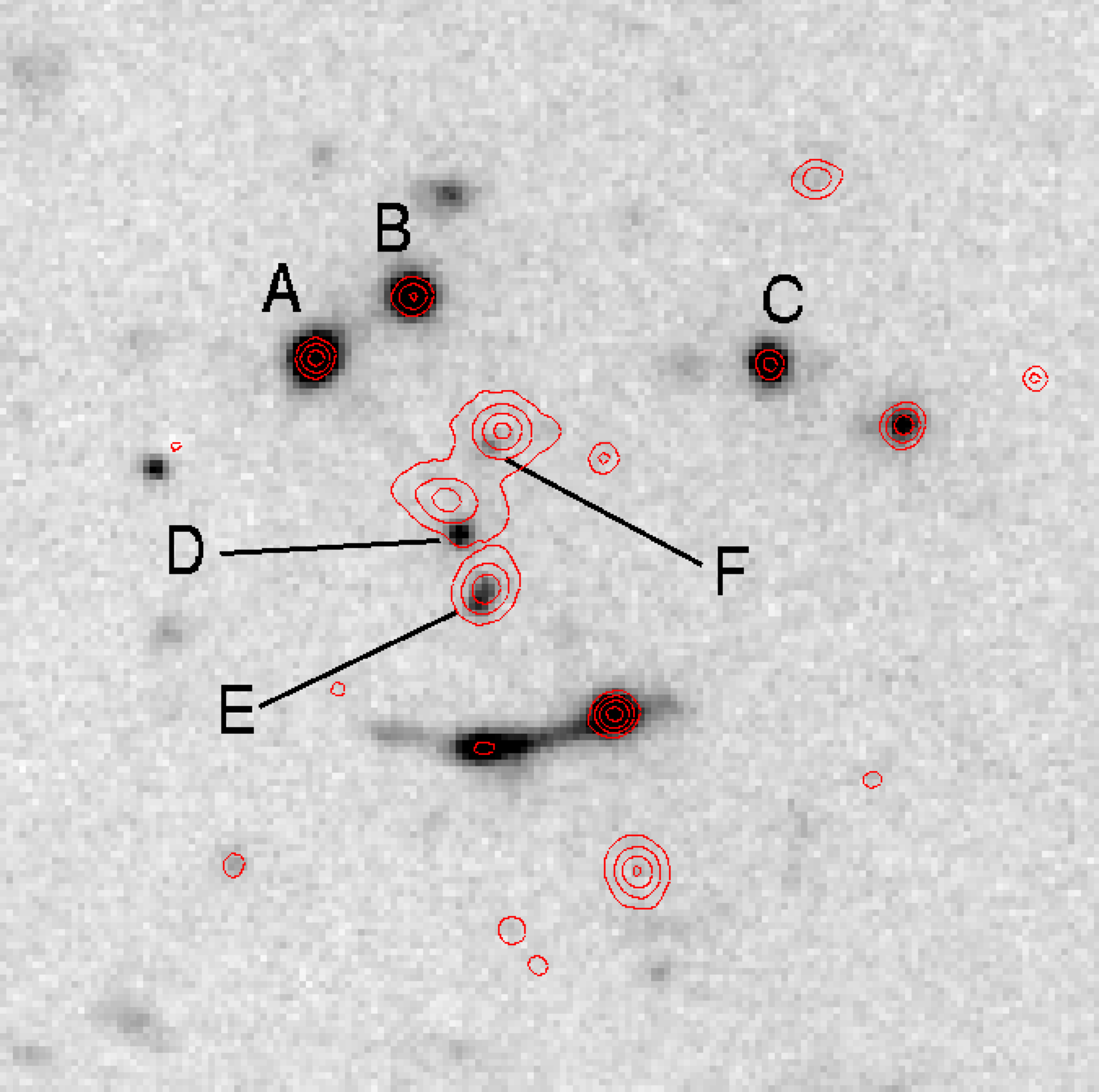}
\caption[.]  {Residuals after subtracting the second-epoch stacked
  $i$-band image from the $g$-band image. Prior to the subtraction,
  the $i$-band image was PSF-matched by convolution with an elliptical
  gaussian kernel and then scaled to match the $g$-band flux of
  galaxies G1-G3. The field of view and orientation is the same as for
  Figure~\ref{fig:Firstepoch}. To indicate the locations of the
  galaxies we also show intensity contours from the $i$-band
  image. Objects with excess blue light in the cluster core are
  candidate quasar images. Images A-E are clearly confirmed by 
  spectroscopy and readily visible in this image subtraction; a
  possible sixth quasar image (F) is also indicated.}
\label{fig:gisub}
\end{figure}

Figure~\ref{fig:gisub} shows a scaled subtraction of the second epoch
$i$-band image from the $g$-band image (see
Table~\ref{tab:MOSCAlog}). The PSF of the $i$-band image was matched
to that of the $g$-band image by convolving it with an elliptical
gaussian kernel. The $i$-band image was then scaled so that the
subtraction optimally removes the light of galaxies G1-G3. Any
remaining flux, shown in the negative in Figure~\ref{fig:gisub}, is
bluer than galaxies G1-G3, and should be considered candidate images
of the quasar.

\begin{figure}
\plotone{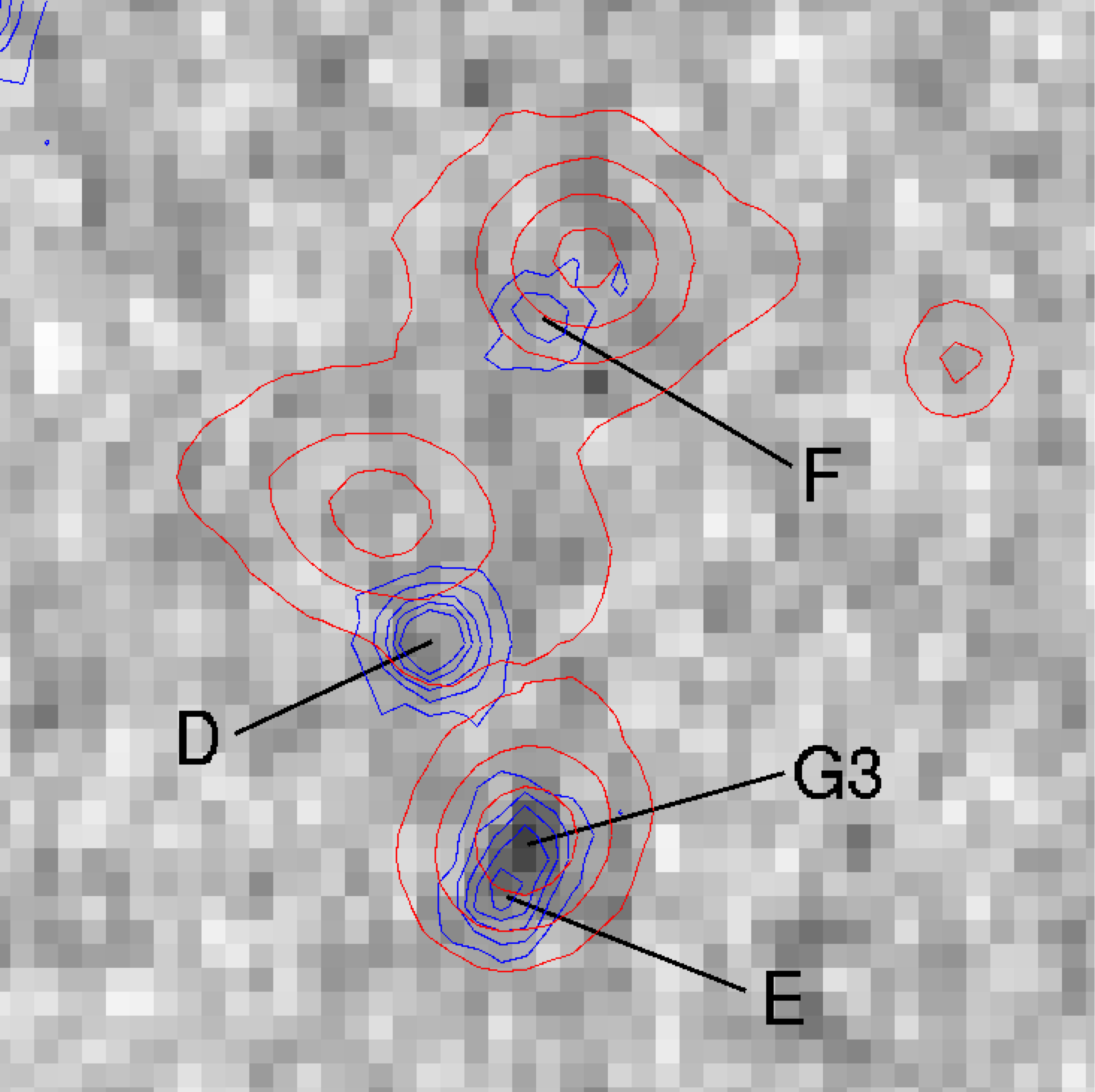}
\caption[.]  {A zoomed view - 4$\times$ finer than
  Figure~\ref{fig:gisub} - of the $u$-band image. Red contours, as
  in Figure~\ref{fig:gisub}, show the second epoch $i$-band
  image. Blue contours show the residual $g$-$i$ shown in greyscale in
  Figure~\ref{fig:gisub}. Note that at the position of image E, the
  residual blue light is extended, and that it partially coincides
  with an excess of $u$-band light that is most prominent at the
  center of galaxy G3. The peak of the excess blue light in $g$-$i$ is
  offset from the center of G3; we interpreted this light as the sum
  of an image of the quasar (E) and a blue core in galaxy G3.}
\label{fig:uzoom}
\end{figure}

In Figure~\ref{fig:gisub}, image D is obvious. A comparison of this
residual image to the residual quasar images A-C shows that it has a
PSF consistent with a point source, as expected. Additionally there is
excess blue flux at the spectroscopically indicated position of image
E; this flux is obviously extended however. Figure~\ref{fig:uzoom},
which shows the $u$-band image with contours from the $g$-$i$
subtraction overlayed, shows the reason for this; the excess blue
light is a combination of light from the core of galaxy G3, and offset
from this, excess blue light attributable to image E.

Finally, in Figure~\ref{fig:gisub} there is also an additional blue
excess source, near galaxy G1, which we label F in
Figures~\ref{fig:gisub} and ~\ref{fig:uzoom}. This source is
consistent with being a point source. One of the slit placements from
the initial spectroscopic observations includes this possible quasar
image, and Figure~\ref{fig:DEFspecs} includes a spectrum extracted
at that location. The extracted spectrum suggests the presence of
Ly$\alpha$ emission; taken together, the imaging and spectroscopic
data suggest that this source is a sixth image of the quasar.

\subsection{Photometry of Quasar Images} 
\label{sec:phot}

To achieve the most accurate photometry of images of the quasar, we
have constructed a model of the lensing cluster core using GALFIT
(Peng et al.\ 2010). The input PSF for each of the models discussed
below was in all cases constructed by fitting bright, isolated and
unsaturated stars using a multi-component Moffat profile using
GALFIT. Typically 3 Moffat components were used; we have explored the
effects of PSF reference star selection and find that it introduces an
uncertainty of less than 0.01 magnitudes for the brighter quasar
images, with no significant biases.

The cluster core model was constructed initially using a stack of the
second epoch $g$, $r$ and $i$ images. Sources were placed initially at
locations of obvious galaxies, and at the six quasar image positions
discussed above.  Quasar images are PSF components, and all galaxies
are S\'{e}rsic profile components. We continued adding components to
the model iteratively at the locations of any significant residuals
revealed by the modeling process, until a final complete model
incorporating all significant sources was achieved. This model was then
translated to individual epoch and filter images by transforming the
image coordinates to match the image to be analyzed, fixing those
coordinates, and then refitting the model component magnitudes, sizes
and shapes to the image to be analyzed.

The modeling process described above reveals that the images of the
quasar include a contribution from the underlying host galaxy, as is
seen in the two other quasars strongly lensed by galaxy clusters
(Inada et al. 2005; Oguri et al. 2012). The data show that the host
galaxy is blue - unlike the red hosts seen in the other two
quasar-cluster lenses - and is most clearly seen in the brightest (and
thus most likely the most magnified) image A. Figure 8 shows residuals
from a GALFIT model for image A in which only a PSF component is used,
and a second model consisting of a PSF component plus a single
S\'{e}rsic component; this second model is a much better fit.

\begin{figure}
\plottwo{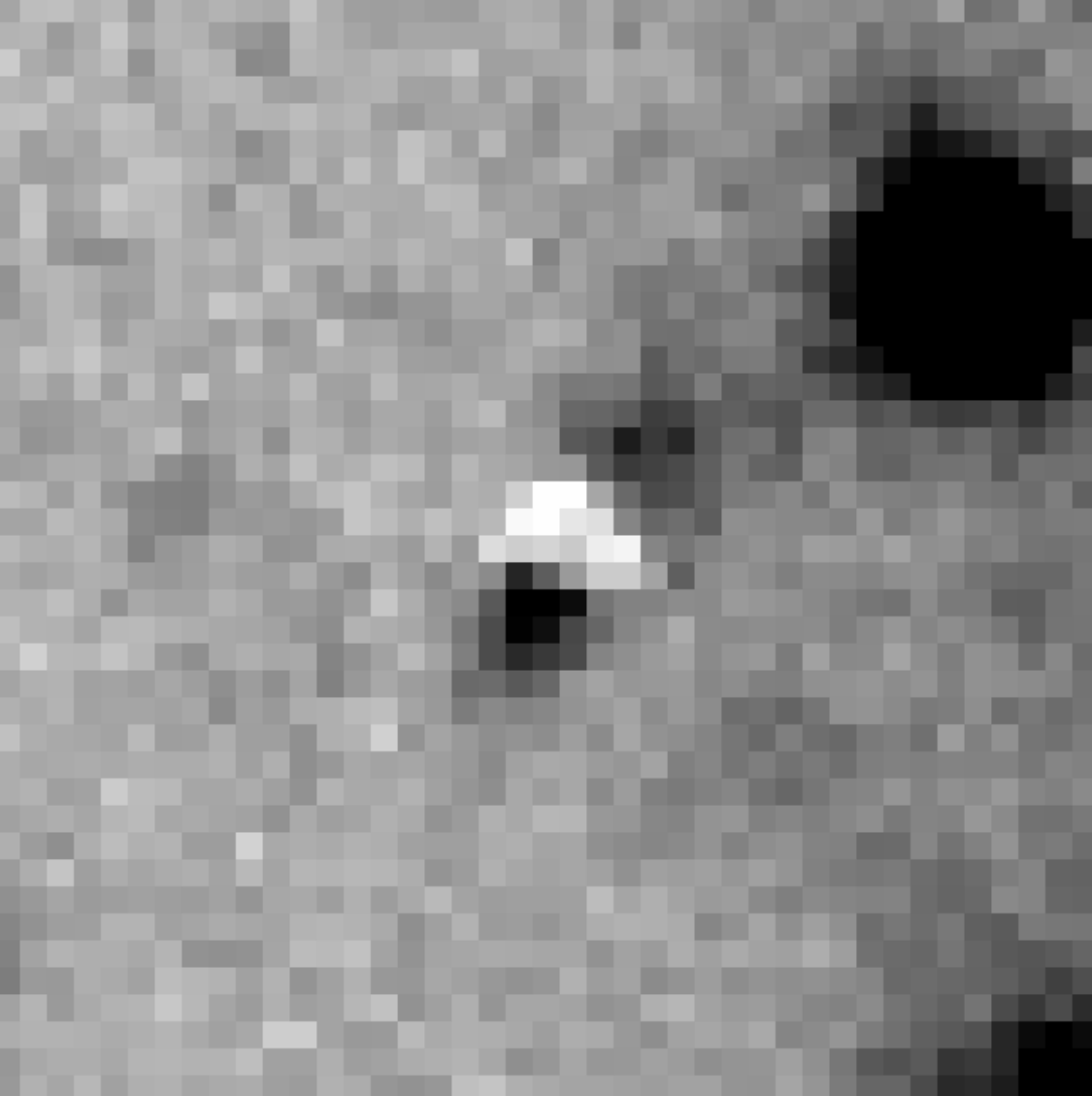}{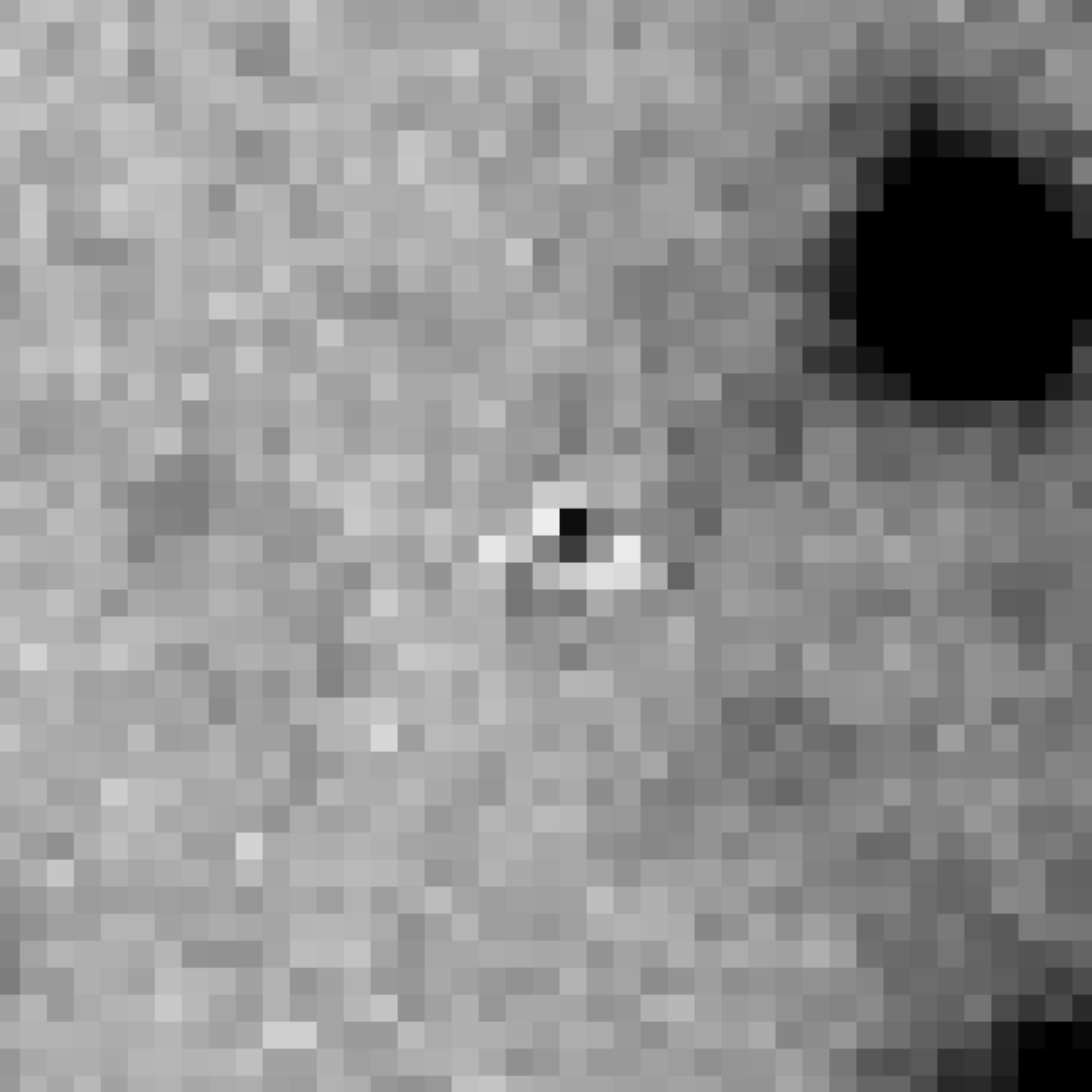}
\caption[.]  {Left Panel: The residuals in a 40$\times$40 pixel region
  from the $g$-band image resulting from a single PSF GALFIT model
  centered on image A. Right Panel: the same region fit by a
  PSF+S\'{e}rsic profile. Note that the PSF-only fit clearly indicates
  the presence of an underlying host galaxy component, and that it is
  oriented tangentially to the cluster center, as expected, and that
  this component is well fit by a single S\'{e}rsic profile.
 }
\label{fig:hostgal}
\end{figure}

In the $g$-band, including the host galaxy makes the fitted quasar
0.086 magnitudes fainter. In our judgement, none of the other images
of the quasar allow for clear modeling of the host galaxy component -
image B is bright enough, but complicated by a second galaxy less than
two arcseconds to the north, and images C-F are insufficiently
magnified. Thus, for images B-D, we proceed by measuring the quasar
brightness using a simple PSF-only fit in GALFIT, and then correct
each measured magnitude fainter by the host galaxy contribution as
measured from image A. Though the quasar and host magnifications are
not expected to be identical across the images, and the quasar/host
ratio will not be completely fixed due to temporal variation of the
quasar (see below), this simple procedure nevertheless avoids
systematic noise that would be introduced by attempting to model the
host galaxy independently in multiple images. For reference, in
addition to the $g$-band above, the correction for the host galaxy in
$r$ and $i$ is measured to be 0.084 and 0.071 magnitudes,
respectively. Within the uncertainties, this correction is the same
across bands - i.e., the host galaxy has a color comparable to the
quasar. For the $u$-band the image is of insufficient depth to model
the host galaxy directly, we adopt the same correction as the
$g$-band. The resulting photometry for images A-D, across four bands
and two epochs, is reported in Table 3. Uncertainties are calculated
by using the full GALFIT model to generate realisations of each
measured frame with appropriate noise, which are then refit. The
measured scatter in the photometry of quasar images, combined in
quadrature with the estimated calibration uncertainty, is
reported. For the brighter images, systematics due to uncertain
treatment of the host galaxy are likely dominant, and not addressable
with these limited data.

Images E and F must be treated with further care. The photometry of
image E is complicated by the blue core of galaxy G3, and image F is
simply very faint, and near the core of the brightest cluster galaxy,
G1. Particularly at redder wavelengths, this makes its measurement
highly uncertain. Furthermore, tests using the full GALFIT model, in which we
re-measured noisy realisations of the full model, also show
potentially strong biases - up to factors of 3 - in the photometry of
image F. To attempt best possible photometry for these we proceeded as
follows.

First, note that if the colors of the various quasar images are the
same, then the relative photometry of residual images in the $g$-$i$
subtraction in Figure~\ref{fig:gisub} can be used to measure the fainter
images. Matched colors are not guaranteed, due both to intrinsic
variability (e.g., Schmidt et al. 2012) coupled with time delays, as
well as variations induced by differential magnification between the
quasar and underlying host galaxy, and possible extinction of
individual images due to intervening galaxies in the complex cluster
core. However, we can validate the likely variability in color between
the images by considering the three brightest and most isolated images
(A-C); these show that the typical scatter about a fixed color is 0.05
in $g$-$i$ and 0.03 in $g$-$r$. This uncertainty is much less than the
systemic biases noted above in attempts at direct measurement, and so
the fixed color assumption is sufficient given these data.

Thus for image F, we use difference images to measure the differential
flux of image F relative to images A-C. Assuming that F has the same
color as these images, the flux of image F is then given simply by the
mean flux of images A-C times the flux ratio in the difference
image. All magnitudes are measured using PSF fitting photometry, for
consistency. The difference image $g$-$i$ provides the best
measurement of the residual fluxes, and is the basis of the
measurements reported in Table 3. As before, we estimate uncertainties
by using the fitted GALFIT model to generate and refit noisy
realisations of the measured image. In the case of image F, the final
reported uncertainty is the scatter in the magnitude offset in the
residual image, combined in quadrature with the color scatter between
the brighter images.

For image E we must further account for the residual flux from the
core of galaxy G3. We proceed as for image F, but additionally use
GALFIT to model a best-fitting S\'{e}rsic profile fixed at the center
of G3 concurrent with the PSF component for image E.

Finally, we note that each frame is directly calibrated to the SDSS
using many hundreds of objects; calibration uncertainties are 0.01
magnitudes at worst.

\begin{deluxetable*}{lcccccc}
\tabletypesize{\footnotesize}
\tablecaption{ Photometry of QSO images \label{tab:QSOphotom}}
\tablehead{
\colhead{Data Frame} & \multicolumn{6}{c}{Quasar Image}\\
\colhead{} & \colhead{A}& \colhead{B}& \colhead{C}& \colhead{D}& \colhead{E}& \colhead{F}
}
\startdata
\hline \\                     
$g$-band 1st epoch   &   21.03$\pm$0.01 & 21.52$\pm$0.01 & 22.13$\pm$0.01 & 23.48$\pm$0.04 & 24.32$\pm$0.09 & 25.21$\pm$0.24 \\
$g$-band 2nd epoch   &   21.15$\pm$0.01 & 21.57$\pm$0.01 & 22.22$\pm$0.01 & 23.74$\pm$0.03 & 24.22$\pm$0.08  & 24.81$\pm$0.07 \\ \hline
$r$-band 1st epoch   &   20.66$\pm$0.01 & 21.13$\pm$0.01 & 21.69$\pm$0.01 & 23.13$\pm$0.11 & 23.92$\pm$0.09 & 24.81$\pm$0.24 \\
$r$-band 2nd epoch   &   20.80$\pm$0.01 & 21.16$\pm$0.01 & 21.79$\pm$0.01 & 23.46$\pm$0.08 & 23.82$\pm$0.08 & 24.41$\pm$0.07 \\ \hline
$i$-band 1st epoch   &   20.82$\pm$0.01 & 21.20$\pm$0.01 & 21.81$\pm$0.01 & 23.46$\pm$0.12 & 24.04$\pm$0.09 & 24.92$\pm$0.24 \\
$i$-band 2nd epoch   &   20.95$\pm$0.01 & 21.31$\pm$0.01 & 21.90$\pm$0.01 & 23.78$\pm$0.08 & 23.96$\pm$ 0.08 & 24.55$\pm$0.07 \\ \hline
$u$-band 2nd epoch   &   22.85$\pm$0.08 & 23.29$\pm$0.09 & 24.18$\pm$0.17 & $>$24.58 &  $>$24.58 & $>$24.58 \\
\tablenotetext{}{Uncertainties are quoted at the 1-$\sigma$
  level. Images D,E,F are not detected in the $u$-band; the 1-$\sigma$
  detection limit in our data is 24.58 mag. When comparing this photometry to the SDSS photometry reported in Table 2, a correction for the quasar host galaxy 
  should be applied to the SDSS photometry, as detailed in the main text.}
\end{deluxetable*}

\subsection{Lensing cluster} 
\label{sec:cluster}

\begin{figure}
\includegraphics[angle=0,scale=.35]{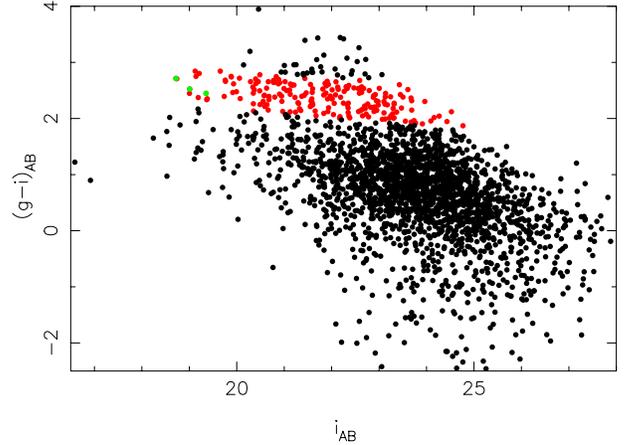}
\caption[.]
{Color-magnitude diagram of 3033 non-stellar objects detected in the MOSCA field containing \QSO . The galaxies G1-3 are indicated by green dots. Other galaxies with $g$-$i$ colors within $\pm 0.35$ magnitudes from the red sequence formed by early-type galaxies in the lensing cluster are indicated by red dots.  }
\label{fig:cmag}
\end{figure}

\begin{figure}
\plotone{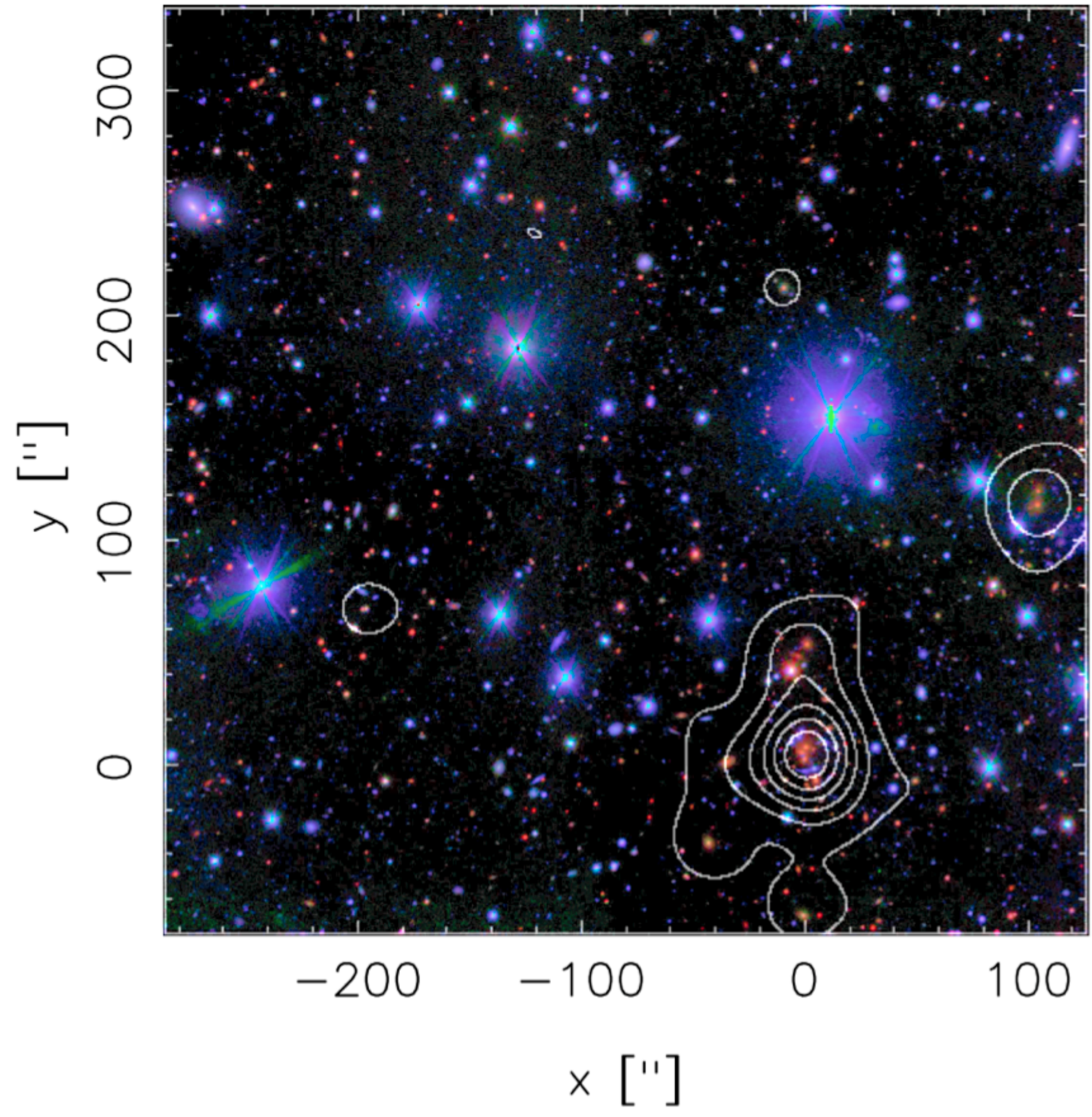}
\caption[color contour plot.]
{A contour plot showing a smoothed map of the light distribution of the red sequence galaxies indicated by the red and green colors in 
Figure~\ref{fig:cmag}. The background figure is a $gri$ composite based on the second-epoch MOSCA images. The coordinates are relative to the position of galaxy G1, which closely coincides with the cluster lens (see \S~\ref{sec:model}). A strong peak in the light distribution is seen at this location.}
\label{fig:lightcont}
\end{figure}

The maximum separation of the quasar images is $15\farcs 1$ (between quasar images A and C) 
clearly indicative of a cluster-sized lensing mass. The SDSS photometric redshift estimates 
of galaxies G1-G3 (see Table~\ref{tab:astrom}) are consistent with all three galaxies being located in a cluster at 
$z \sim 0.5$. As part of the observations described in \S~\ref{sec:ABCspec} and \S~\ref{sec:DEF} above, all three galaxies were targeted
spectroscopically. Redshift estimates were obtained through cross-correlation with the elliptical galaxy template of Kinney et al.\ (1996), 
yielding the redshift estimates listed in Table~\ref{tab:galaxies}. The redshifts are consistent within the measured uncertainties, and  
all three galaxies G1-G3 clearly belong to the same cluster at a redshift consistent with the SDSS photometric redshift estimates. 
To further investigate the properties of the lensing cluster, we applied the SExtractor software (Bertin \& Arnouts 1996) for galaxy photometry 
to the stacked second-epoch MOSCA $gri$ images. Prior to obtaining the photometric measurements, the stacked $r$- and $i$-band images 
were smoothed to match the resolution of the $g$-band image. Object detection was initially performed on the stacked $g$-band image, 
and SExtractor was subsequently run in dual-image mode on the stacked images in the other passbands, using the $g$-band image 
as a reference for object detection. The observed magnitudes were subsequently converted to AB magnitudes in the SDSS system using  
objects in the MOSCA field with catalogued SDSS photometry. Total magnitudes were determined using the SExtractor \texttt{MAG\_AUTO} parameter, 
and colors were determined from aperture magnitudes (\texttt{MAG\_APER}) within a $2\arcsec$ diameter aperture.    
A $g$-$i$ vs.\ $i$ color-magnitude diagram of non-stellar objects, selected based on their location in a magnitude - $\mu_{\rm max}$ diagram 
(where $\mu_{\rm max}$ is the central surface brightness of objects), is shown in Figure~\ref{fig:cmag}. 
A red sequence is evident at $g$-$i \sim 2.5$, similar to the $g$-$i$ colors of the three galaxies G1-G3 (see Table~\ref{tab:galaxies}). 
Figure~\ref{fig:lightcont} shows a smoothed map of the light distribution of red sequence galaxies over the observed MOSCA field.  
A strong peak in the luminosity distribution is located at the position of galaxies G1-G3.           

\begin{deluxetable*}{lcccc}
\tabletypesize{\footnotesize}
\tablecaption{ Photometry and redshift measurements for galaxies in the cluster core.\label{tab:galaxies}}
\tablehead{
\colhead{Galaxy} & \colhead{i} &  \colhead{g$-$i} & \colhead{r$-$i} & \colhead{Redshift} \\
      & \colhead{ } & \colhead{($2\farcs 0$ aperture)  } & \colhead{($2\farcs 0$ aperture) } &  \colhead{ } 
}
\startdata
\hline \\                     
G1         & 18.74 & 2.71 & 0.97 & $0.493 \pm 0.013$ \\
G2         & 18.97 & 2.53 & 0.96 & $0.496 \pm 0.012$ \\
G3         & 19.37 & 2.45 & 0.93 & $0.489 \pm 0.013$ \\
\end{deluxetable*}

\subsection{Giant arc} 
\label{sec:arc}

The identification of \QSO \, as a candidate gravitational lens system was partially due to recognition of the giant arc just 
south of the cluster center. The extent of the arc in our deepest stacked $g$-band image (see Figure~\ref{fig:Slitpos}) is $10\arcsec$.   
Our SExtractor photometry (see \S~\ref{sec:cluster}) of the arc yields $g=21.57$, and $u$-$g = 0.87$, $g$-$r = 0.21$, $g$-$i = 0.41$,  
where the colors are measured in $2\arcsec$ diameter apertures centered on the brightest central portion of the arc. 
We note that the quoted magnitude of the arc can be considered a lower limit on its brightness, given the very extended mophology of the arc. 
Although SEXtractor attempts to treat the arc and the WD as separate objects, a more careful modeling of the shape of the arc (including PSF modeling and removal of the superposed white dwarf point source) 
would be required to obtain an accurate total magnitude.
The spectrum of the arc shows a single emission line at $\lambda_{obs} = \sim 4015${\AA}, which we interpret as Lyman-alpha at $z=2.30$. 
The alternative interpretation of [O II] (3727 {\AA}) at $z=0.077$ would put the arc in front of the cluster, which is obviously inconsistent 
with the lensed morphology of the arc. The lens model presented in the following section does not predict a counter-image of the arc.

\section{Lens model} 
\label{sec:model}

We compute a parametric lens model using the publicly available software \texttt{Lenstool} (Jullo et al. 2007), through Markov Chain Monte Carlo minimization in the image plane. The lens is modeled with several components: a cluster halo, BCG, and cluster galaxies.
All the components are represented by a pseudo-isothermal elliptical mass distributions (PIEMD\footnote{This profile is formally the same as dual Pseudo Isothermal Elliptical Mass Distribution (dPIE, see El{\'{\i}}asd{\'o}ttir et al. 2007).}; Limosin et al. 2005, El{\'{\i}}asd{\'o}ttir et al. 2007), parametrized by its position $x$, $y$; a fiducial velocity dispersion $\sigma$; a core radius $r_{core}$; a scale radius $r_s$; ellipticity $e=(a^2-b^2)/(a^2+b^2)$, where $a$ and $b$ are the semi major and semi minor axes, respectively; and a position angle $\theta$. All the parameters of the cluster halo distribution are allowed to vary within priors, except for $r_s$, which is not constrained by the lensed images.

Cluster members are selected based on their $g$-$i$ color in a color-magnitude diagram, with respect to the cluster red-sequence (see Figure~\ref{fig:cmag}, and their position, ellipticity and position angle are fixed at their observed values, as measured with SExtractor. The parameters $\sigma$ and $r_s$ are determined through scaling relations (see Limousin et al.\ 2005 for a description of the scaling relations).
Three galaxies, G1, G2 and G3, are in close proximity to three lensed images of the quasar, and thus affect their positions more directly than other cluster members. To account for the intrinsic scatter in the scaling relations, we allow some of the parameters of these three galaxies to be modeled separately from the rest of the cluster galaxies. 

As constraints, we use the redshift and positions of the images of the quasar and of the lensed galaxy at $z=2.30$.
We have a total of 12 free parameters: 6 free parameters for the cluster halo, and 2 free parameters for each of the galaxies, and 14 constraints. The best fit parameters are listed in Table~\ref{t.parameters}. The uncertainties were estimated through the MCMC sampling of the parameter space. For parameters that were not well-constrained, we list the priors that were set in the minimization.
To account for systematics due to misidentification of the sixth image, we also compute a model assuming only five images. The two families of models are similar, although we note that G1 is poorly constrained in the model without image F.

The image-plane RMS scatter of the quasar images is 0.14'' in the 5-image model and 0.18'' in the 6-image model; the RMS of the blue arc is somewhat larger (0.26'',0.38'', respectively), due to the difficulty of measuring its positional constraints, and the larger weight given to the quasar images in the modeling process. 

Both models successfully predict the locations of all the observed
images (either five or six, depending on the model constraints), and
do not predict additional images, with the exception of demagnified
central images that cannot be detected in our data: the model with six
images predicts a demagnified seventh image close to the center of
galaxy G2. This is expected in this strong lensing configuration. Our
imaging data do not reveal any hint of this seventh image, though a
positive identification might be possible with the superior spatial
resolution of {\it HST}. A careful re-examination of the spectroscopic
data for G2 shown in Figure~\ref{fig:2Dspec} and plotted in
Figure~\ref{fig:DEFspecs} may indicate a weak signature of the
Lyman-alpha line from the seventh image (G), but this feature is
almost at the level of the noise, and more sensitive spectroscopy with
8-10m class telescopes will be required for a definite detection.

Table \ref{t.mag} lists the model-predicted magnifications and time
delays for the two model families. Uncertainties are computed from a
suite of models, drawn from the MCMC chain, that represent the
1-$\sigma$ uncertainty in the parameter space. All time delays are
given relative to image A.  The best-fit models predict short time
delays of order of 100 days between the close pair (A-B), and between
the central images (D-E-F). Image C precedes A and B by 1400 days, and
D-E-F follow A and B by $700-1000$ days.  In the 6-image model, 66\%
of the models, as well as the best-fit model, predict the order of the
images to be C-B-A-F-E-D. All the models predict that image C is the
first one to occur, followed by B-A (in 93\% of the models) or A-B
(7\%); followed by F-E-D (72\%) F-D-E (16\%) or D-F-E (12\%).  In the
5-image model, 62\% of the models, as well as the best-fit model,
predict C-B-A-E-D; here too, all the models predict that C is the
first image to occur, followed by B-A (77\%) or A-B (23\%), and then
E-D (81\%) or D-E (19\%). 
    
\begin{figure}
\includegraphics[angle=0,scale=.425]{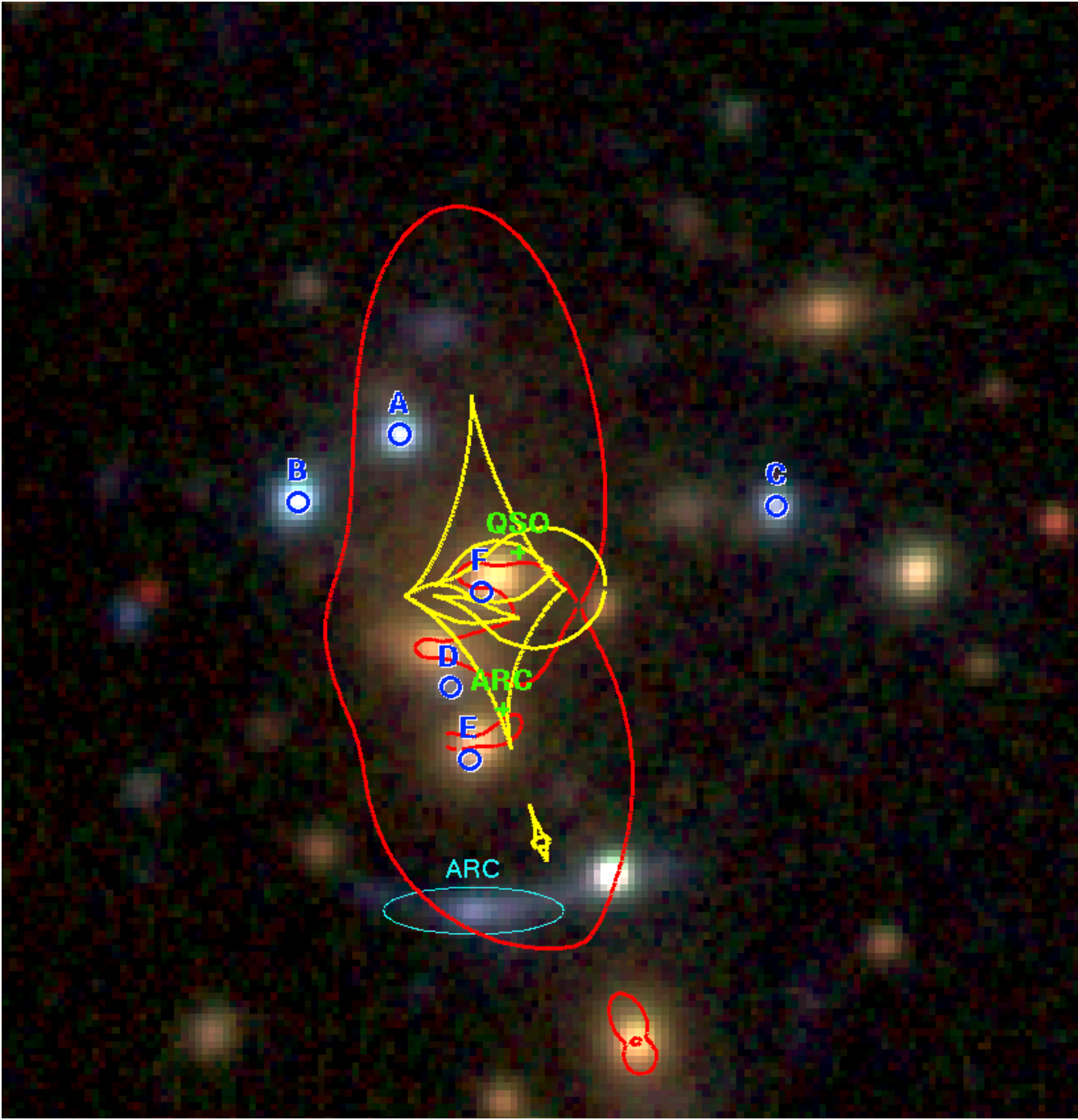}
\caption[.]
{Caustics, critical lines, and image positions from the best-fit model, superposed on a $gri$ color image of the \QSO \, lens system. The critical curves for a source 
at the quasar redshift ($z=2.82$) are indicated in red. The caustic curves in the source plane are shown as yellow lines, 
and the predicted source locations of the quasar and arc with respect to these curves are indicated by green crosses. The observed locations of the six quasar images A-F are indicated by blue circles.   
}
\label{fig:Model}
\end{figure}

\begin{deluxetable*}{lccccccc} 
 \tablecolumns{8} 
\tablecaption{Best-fit lens model parameters  \label{t.parameters}} 
\tablehead{\colhead{Halo }   & 
            \colhead{RA}     & 
            \colhead{Dec}    & 
            \colhead{$e$}    & 
            \colhead{$\theta$}       & 
            \colhead{$r_{\rm core}$} &  
            \colhead{$r_{\rm cut}$}  &  
            \colhead{$\sigma_0$}\\ 
            \colhead{(PIEMD)}   & 
            \colhead{($\arcsec$)}     & 
            \colhead{($\arcsec$)}     & 
            \colhead{}    & 
            \colhead{(deg)}       & 
            \colhead{(kpc)} &  
            \colhead{(kpc)}  &  
            \colhead{(km s$^{-1}$)}  } 
\startdata 
\cutinhead{Model 1: constraints from 5 QSO images. RMS$_{\rm QSO}$=0.14''}
Cluster  & \Px        & \Py         & \Pe    & \Ptheta    & \Prc   & [2000] & \Psigma  \\ 
G1       & [0.0]      & [0.0]       & [0.26] & [3.8]         & \PrcA & [47]     & \PsigmaA  \\ 
G2    & [-0.596] &[-5.220] & [0.21]  &[75.5]       & \PrcB & [24]     & \PsigmaB \\ 
G3    & [-1.762] &[-2.308] &[0.41] &[-18.3]       &\PrcC  &[20]     & \PsigmaC \\ 
L* galaxy  & \nodata & \nodata & \nodata & \nodata &  [0.15]  &     [20]&  [100]  \\ 
\cutinhead{Model 2: constraints from 6 QSO images.  RMS$_{\rm QSO}$=0.18''}
Cluster  & \Pxsix        & \Pysix         & \Pesix    & \Pthetasix    & \Prcsix   & [2000] & \Psigmasix  \\ 
G1       & [0.0]      & [0.0]       & [0.26] & [3.8]         & \PrcAsix & [47]     & \PsigmaAsix  \\ 
G2    & [-0.596] &[-5.220] & [0.21]  &[75.5]       & \PrcBsix & [24]     & \PsigmaBsix \\ 
G3    & [-1.762] &[-2.308] &[0.41] &[-18.3]       &\PrcCsix  &[20]     & \PsigmaCsix \\ 
L* galaxy  & \nodata & \nodata & \nodata & \nodata &  [0.15]  &     [20]&  [100]  \\ 
\enddata 
 \tablecomments{All coordinates are measured in arcseconds relative to the center of G1, at [RA, Dec]=[335.53573, 27.75986]. The ellipticity is expressed as $e=(a^2-b^2)/(a^2+b^2)$. $\theta$ is measured north of West. Error bars correspond to 1 $\sigma$ confidence level as inferred from the MCMC optimization. Values in square brackets are for parameters that are not optimized. Errors in square brackets represent the lower and upper limits that were set as prior in the optimization process, for parameters that were not well-constrained by the data. 
The location and the ellipticity of the matter clumps associated with the  cluster galaxies were kept fixed according to their light distribution.}
\end{deluxetable*}

\begin{deluxetable*}{lccccccc} 
 \tablecolumns{8} 
\tablecaption{Model-predicted magnifications and time delays\label{t.mag}} 
\tablehead{\colhead{Image }   & 
            \colhead{Magnification (1)}     & 
            \colhead{Time delay (1) [days]}    & 
            \colhead{Magnification (2)}     & 
            \colhead{Time delay (2) [days]}    &   
} 
\startdata 
A  & \magA & \timeA &\magAsix &\timeAsix  \\ 
B  & \magB & \timeB &\magBsix &\timeBsix  \\ 
C  & \magC & \timeC &\magCsix &\timeCsix  \\ 
D  & \magD & \timeD &\magDsix &\timeDsix  \\ 
E  & \magE & \timeE &\magEsix &\timeEsix  \\ 
F  & \nodata & \nodata &\magFsix &\timeFsix  \\ 
\enddata 
 \tablecomments{Model (1): constraints from 5 QSO images.   Model (2): constraints from 6 QSO images. Time delay is given in days, relative to image A.}
\end{deluxetable*}

\section{Summary and discussion} 
\label{sec:discuss}

\subsection{Previous observations and predictions of high image multiplicities} 
\label{sec:multiplicity}

We have identified \QSO \, as a gravitational lens system producing six observed images of the same quasar. This is the third known 
cluster-scale lens and the only such lens producing more than five observed images. The only previously known six-image AGN lens 
is B1359$+$154 (Rusin et al.\ 2001), where a radio source at $z_s = 3.235$ is being lensed by a compact group of three galaxies at $z_l \sim 1$. 
However, the quasar image separations we observe in \QSO \, are an order of magnitude larger than in B1359$+$154, and the lensed images are 
$\sim 3$ magnitudes brighter at optical wavelengths, making six images detectable in ground-based imaging of \QSO . 
The B1359$+$154 lensed image configuration is similar to what we find in \QSO , with the three fainter images located in close proximity to the 
three galaxies situated within the Einstein radius of the lens system, and three brighter images located further away from these galaxies. 
However, the lens model of B1359$+$154 presented by Rusin et al.\ (2001) does not allow a significant dark matter halo associated with the
galaxy group in addition to the three halos corresponding to the three individual galaxies. In contrast, in our model the cluster halo 
component is the dominant contributor to the observed image splitting of \QSO . 
 
Since cluster-scale quasar lenses and image multiplicities $>5$ are both rare, we would expect a lensing system such as 
\QSO \, to be extremely rare. There are no appropriate quantitative predictions for the 
occurrence of this type of lens systems available in the literature. Evans \& Witt (2001) calculate the incidence of sextuplet 
and octuplet image systems (with demagnified seventh and ninth images, respectively) 
and estimate that these constitute $\sim 1\%$ of the population of quasars multiply imaged by galaxy-scale lenses.  
This is consistent with the single known sextuplet case of B1359$+$154 among the $\sim 100$ known galaxy-scale quasar lenses. 
For the larger image separations produced by cluster lenses, published 
studies of the probabilities of different lensing multiplicities (Oguri \& Keeton 2004; Li et al.\ 2007; Minor \& Kaplinghat 2008) 
only include double, quadruple (with demagnified third and fifth images, respectively) and naked cusp configurations. 
Li et al.\ (2007) state that they do not expect a significant fraction of sources to have image numbers exceeding 5.

\subsection{Comparing the lens and source redshifts to predictions} 
\label{sec:redshifts}

Both the quasar source redshift and cluster lens redshift of \QSO \, are comparable to the two other cluster quasar lenses and in good 
agreement with theoretical expectations: The observed lens redshift in \QSO \, is $z=0.49$, which is similar, but slightly lower than 
the cluster lenses in \QSOLone \, ($z=0.68$) and \QSOLtwo \, ($z=0.58$). This redshift is very close to the peak (at $z \simeq 0.5$) of the predicted lens redshift distribution for photometrically selected wide-separation quasars in the SDSS (Li et al.\ 2007). 
The observed quasar source redshift in \QSO \, is $z=2.82$, slightly higher than the sources in \QSOLone \, ($z=1.73$) 
and \QSOLtwo \, ($z=2.20$). These three redshifts are all near the peak (around $z=2$) of the predicted source redshift distribution for 
photometrically selected wide-separation quasars in the SDSS (Li et al.\ 2007).

\subsection{Prospects for time delay measurements} 
\label{sec:delta_t}

The photometry reported in \S~\ref{sec:phot} shows significant variability in the quasar, demonstrating the feasibility of making time 
delay measurements in \QSO \, from monitoring campaigns using ground-based 2-4m class telescopes. We have recently initiated a monitoring 
program for \QSO \, 
at the NOT for this purpose. The predicted time delays from our lens model (see \S~\ref{sec:model}) indicate that three independent 
time delays $\Delta t_{AB}$ ($\equiv \Delta t_{B} - \Delta t_{A}$), $\Delta t_{ED}$ and $\Delta t_{FE}$ in the system are all of order 100 days. 
This suggests that several time delay measurements could, in the best case, be obtained from a single season of monitoring. 
The predicted time delays between image C and the close pair (A-B) and between (A-B) and the central images (D-E-F) 
are of order 1400 days and 700-1000 days, respectively. These values are similar to the two longest predicted time delays for \QSOLone , of 1218 and 1674 days, respectively (Oguri 2010) and the measured longest time delay in \QSOLtwo \, of $\Delta t_{AB} = (744 \pm 10)$~days (Fohlmeister et al.\ 2012).     

\subsection{Future work} 
\label{sec:future}

Morphological, photometric and spectroscopic evidence presented in this paper strongly suggest the presence of a sixth 
(F) image of the quasar located closest to the lensing mass center. The odd number theorem of strong gravitational lensing theory 
will then require the presence of a seventh (G) image, for which our model predicts a location close to the center of cluster galaxy G2.
In the case of galaxy-scale lenses, the faintest image of a multiply lensed quasar is typically highly demagnified and hence unobservable 
in the vast majority of cases. However, the shallower central slopes of the mass density profiles of clusters 
increases the chance of observing the final odd image, as exemplified by the identification of the demagnified fifth central 
image in \QSOLone \, using {\it HST} imaging (Inada et al.\ 2005, 2008).   
   
While the data presented in this paper do not provide conclusive evidence for the presence of the seventh image, 
there is a hint of Lyman-alpha emission from this image in the spectrum of G2. If this feature is indeed emission from the quasar, 
a firm confirmation would be possible through spectroscopic observations with 8-10m class telescopes.  

While time delay measurements would provide important additional modelling constraints on the \QSO \, lens system, follow-up 
HST imaging will also be crucial to improve the modeling constraints from the observed locations of the quasar images and hopefully 
provide morphological and photometric confirmation of the predicted seventh (G) image close to the center of galaxy G2. Further constraints 
on the lens system could also come from spatially resolved, multiply imaged structures within the giant arc and from the identification of additional fainter, 
multiply imaged, galaxies as revealed by the HST observations of \QSOLone \, (Sharon et al.\ 2005) and \QSOLtwo \, (Oguri et al.\ 2012).

\subsection{Broader significance of this discovery} 
\label{sec:significance}

The significance of the discovery of \QSO \, is twofold: Firstly, the presence of six quasar images (and potential future confirmation 
of a seventh image), spread over a range of clustercentric radii, with good prospects for future time delay measurements, 
coupled with the presence of an additional lensed source producing a bright giant arc, can provide unique detailed measurements of the 
matter distribution in the core of a cluster. Such measurements provide a stringent test of the scenario of structure formation, including the 
properties of dark matter in cluster cores.   

Secondly, adding our new discovery of \QSO \, to the previous detections of 
\QSOLone \, and \QSOLtwo \, makes a significant contribution towards establishing quasars strongly lensed by galaxy clusters 
as a cosmological tool. The probability of such rare lenses depends very sensitively on cosmological parameters, in particular 
the normalization of the matter power spectrum $\sigma_8$ and the matter density $\Omega_M$ (e.g., Li et al.\ 2007). 
Even though no directly applicable predictions can be found in the literature, the occurrence of lens systems with image separations 
and image multiplicities as high or higher than \QSO \, must necessarily be exceptionally rare, and this could conceivably be the 
the only such system, provided that the current $\Lambda$CDM paradigm is correct.  


\acknowledgments 

HD acknowledges support from the Research Council of Norway. MDG
thanks the Research Corporation for support of this work through a
Cottrell Scholars award. KS acknowledges support from the University
of Michigan's President's Postdoctoral Fellowship. JPUF acknowledges
support from the ERC-StG grant EGGS-278202. The Dark Cosmology Centre
is funded by the DNRF.  We thank the NOT staff, in particular our
support astronomer Stefan Geier, for excellent support, including
accommodating a last-minute instrument change which allowed the
spectroscopic confirmation of the D image and discovery of the E image
of the quasar.

\end{document}